\documentclass[aog]{igs}

\usepackage[utf8]{inputenc}

\usepackage{todonotes}

\usepackage{mathabx}
\usepackage{graphicx}
\usepackage{siunitx}
\usepackage{comment}
\usepackage{xcolor}

\jourvolume{}
\jourissue{}
\jourpubyear{2026}

\begin{document}

\title[BigRAID performance]{Drilling for the Radio Neutrino Observatory in Greenland (RNO-G): field performance of BigRAID}

\author[Tosi et al.]{Jason Chan,$^1$
  Mirko Kugelmeier,$^1$ 
  Delia Tosi,$^1$
  Emma Fisher,$^2$ 
  Chris Kerr,$^2$
  Scott Polfrey,$^2$
  Julius Rix,$^2$
 James Veale,$^2$
 the RNO-G Collaboration$^3$}  

\affiliation{%
$^1$University of Wisconsin-Madison, WI, US \\
$^2$ British Antarctic Survey, Cambridge, UK \\
$^3$ https://rnog-data.zeuthen.desy.de/authorlist date: June 1 2026 \\
  Correspondence: Delia Tosi 
  \email{delia.tosi@wisc.edu},
  \email{authors@rno-g.org}}

\begin{frontmatter}

\maketitle

\begin{abstract}
The Radio Neutrino Observatory in Greenland (RNO-G) is designed to detect ultra-high-energy neutrinos by capturing the radio signals generated when these rare particles interact with glacial ice. Located near Summit Station on the Greenland ice sheet, RNO-G will ultimately consist of 35 autonomous stations, each spaced 1.25 km apart and equipped with arrays of in-ice antennas. A key component of each station is a set of three subsurface strings of antennas, which require boreholes that are approximately 100 meters deep and more than 285 mm wide in diameter. Construction of the observatory began in 2021, and eight stations have been deployed to date. To meet the drilling requirements, the project utilizes the BigRAID system. Developed through a collaboration between the University of Wisconsin-Madison, the British Antarctic Survey (BAS), and the RNO-G Collaboration, BigRAID is being continually refined to improve its reliability, enhance drilling speed, and incorporate advanced automation and control software. This paper will review recent upgrades to the drill and assess its performance in the field through the four seasons of drilling.
\end{abstract}

\end{frontmatter}

\section{Introduction}

Neutrinos are nearly massless, electrically neutral elementary particles that interact only through the weak nuclear force and gravity. Because they interact so rarely with matter, trillions of neutrinos pass through our bodies every second without leaving a trace. They are produced in a wide variety of astrophysical environments, ranging from the Sun and supernovae to black holes and cosmic-ray interactions. Their weak interaction makes them extremely difficult to detect, but also allows them to travel across the universe undeflected and unabsorbed, carrying direct information about the most energetic and distant cosmic phenomena.

Searches for ultra-high-energy (UHE) neutrinos focus on neutrinos with energies far beyond those produced in terrestrial accelerators, typically above a petaelectronvolt (PeV). These particles are expected to originate from extreme astrophysical sources such as active galactic nuclei, gamma-ray bursts, or from interactions of ultra-high-energy cosmic rays with the cosmic microwave background (the so-called cosmogenic neutrinos). Detecting UHE neutrinos would help identify the sources of cosmic rays and probe particle interactions at energies unreachable from Earth.

\begin{figure*}[h]
   \centering
   \includegraphics[width=1\linewidth]{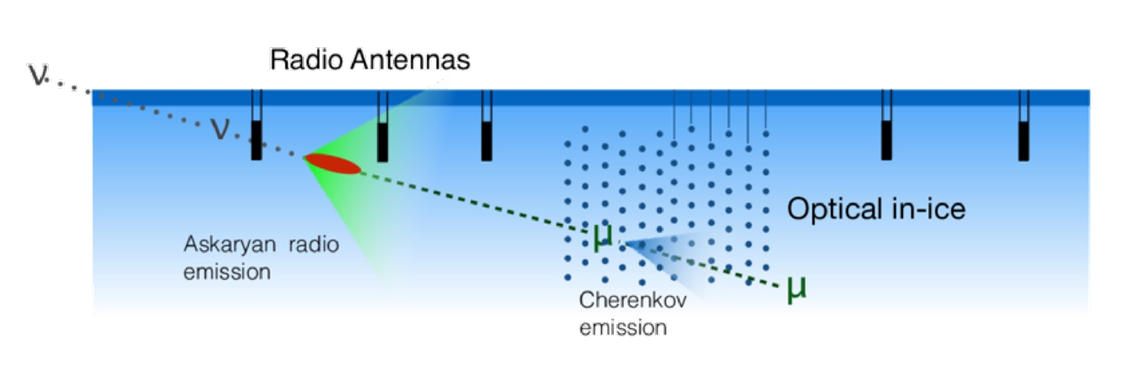}
   \caption{Diagram showing the Askaryan effect. A high energy neutrino interacting in the ice will produce radio waves that can be detected and reconstructed by an array of antennas installed in the ice.} 
   \label{fig:askaryan}
\end{figure*}

Experiments such as the Radio Neutrino Observatory in Greenland (RNO-G) \citep{RNO-G:seven, RNO-G:2020rmc} use large volumes of polar ice as a detection medium. When a UHE neutrino interacts in the ice, it produces a particle cascade that emits a short radio pulse through the Askaryan effect \citep{ANITA:2006nif}. Arrays of buried radio antennas can detect these signals over kilometer-scale distances (Fig.\,\ref{fig:askaryan}), making radio detection a promising technique for observing the extremely rare highest-energy neutrinos \citep{Vieregg:2015baa}.

\begin{figure*}[h]
    \centering
    \includegraphics[width=1\linewidth]{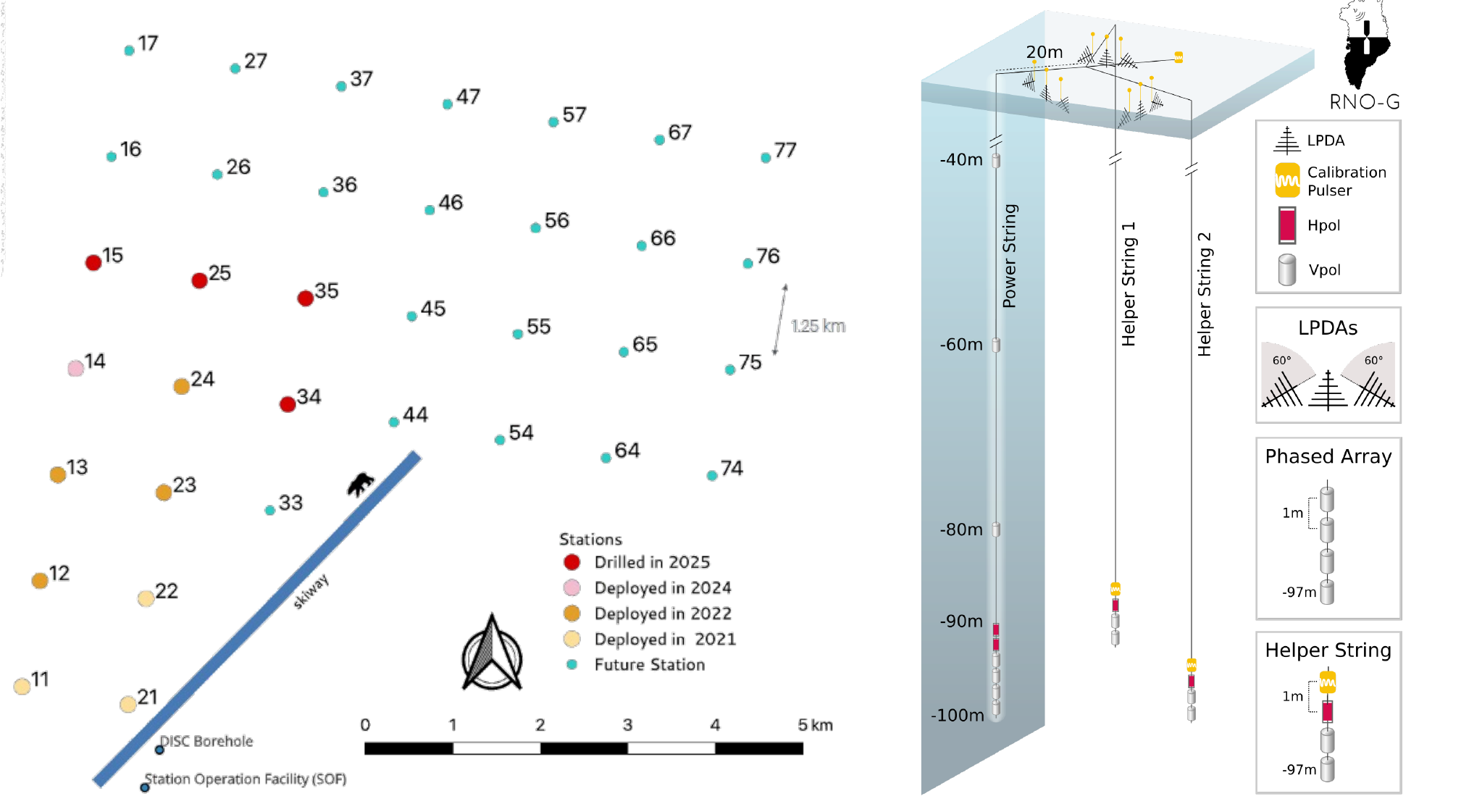}
    \caption{Left: Top view of the RNO-G array as built and planned as of September 2025. Each station (shown with a dot) is marked according to the year it was installed. Right: Diagram of a station, depicting three 100\,m deep holes. One hole is filled with a highly packed string of antennas (which provides sensitivity) and two holes are filled with sparser antennas (for better reconstruction) and a pulser (for calibration). Antennas are vertically polarized (Vpol), horizontally polarized (Hpol) and Log-Periodic Dipole Array (LPDA). More instrumentation including a data acquisition and communication box, solar panels and a wind turbine are deployed at the surface.} 
    \label{fig:RNO-G_schematic}
\end{figure*}

By the time RNO-G will be complete in its first phase, it will feature 35 stations (Fig.\,\ref{fig:RNO-G_schematic}), each including 3 strings of antennas deployed to 100\,m depth. Power is generated by solar panels and wind turbines and communication is implemented through a Long Term Evolution (LTE) network. For a full description, see \citep{RNO-G:seven}.  

\section{Drilling operations} 
\label{drilling_ops}
In 2021, the Radio Neutrino Observatory Greenland (RNO-G) adopted BigRAID as its primary drilling system. BigRAID is a specialized electromechanical drill that was being developed by the British Antarctic Survey (BAS) as a rapid-access firn drill, based on their Rapid Access Isotope Drill (RAID) \citep{Rix:2019}. Given the limited logistical infrastructure at Summit Station and the short field seasons in Greenland, RNO-G required a lightweight and fuel-efficient drilling system capable of repeatedly producing 100\,m-deep boreholes at remote sites with a small field crew and minimal infrastructure, while providing a borehole diameter as large as practical. At the same time, maximizing the borehole diameter within these operational constraints was highly desirable for deployment of the in-ice instrumentation. BigRAID was developed to meet these requirements. Two relevant alternatives considered were hot-water drilling and continuous-auger systems.

\begin{itemize}
\item \textbf{Hot-water drilling systems}:
Hot-water drills, such as the system used for the \emph{Askaryan Radio Array} (ARA) \citep{ALLISON2012457}, can produce large-diameter boreholes but require substantial thermal and electrical power, as well as water-handling, pumping, and support infrastructure. They also generally require a larger field crew. Furthermore, hot-water drilling produces a water-filled borehole, and water can be lost into the porous firn, requiring additional measures before instrumentation can be deployed. 

\item \textbf{Continuous-auger systems}: 
Continuous-auger systems provide another mechanical approach to drilling through firn and ice. In these systems, a helical auger is continuously rotated and advanced into the firn, transporting the cuttings to the surface along the flights. The \emph{Agile Sub-Ice Geological} (ASIG) Drill, developed by the U.S. Ice Drilling Program \citep{Kuhl_2021}, was considered as a possible RNO-G drilling system. The ASIG system was developed for a substantially different application, including deep access and subglacial sampling, and consequently employs a larger and heavier surface system. In addition, ASIG can produce a 146 mm (5.75 in) pilot hole, which, at the time of the instrumentation design, was considered the lower acceptable limit for the RNO-G antennas; however, a larger borehole diameter was preferred, particularly for the design of horizontally polarized antennas. 

\end{itemize}

Unlike traditional coring drills, which are designed to extract continuous ice cores, BigRAID cuts the ice into small chips. The chips are collected within the drill sonde during each drilling run and are ejected and dispersed at the surface when the sonde is returned. This approach enables rapid penetration while producing dry, unobstructed boreholes approximately 285\,mm in diameter. A two-person team can complete a 100\,m borehole in approximately 12--15 h, with a turnaround time of less than two days between successive holes.

A typical drilling season at Summit Station is divided into two operational periods (shifts) of approximately 5 to 6 weeks each, staffed by separate teams. The first shift, typically deployed from mid-May to mid-June, consists of three people, while the second shift, from early June to mid or late July, is usually staffed with two. Each team includes at least one member with extensive drilling experience from previous seasons to ensure operational continuity and knowledge transfer.

At the end of each field season, part of the drilling infrastructure, including the sleds, generators, tower and frame, and a large fraction of the tooling is bermed at Summit Station for overwinter storage. Other components are returned to the United States or United Kingdom for maintenance, repairs, upgrades, and testing.

The first shift is primarily focused on reassembling and commissioning the drill system, a process that typically requires up to two weeks. This period includes the integration and validation of hardware and software upgrades, system troubleshooting, and operational testing. Once commissioning is complete, the team generally drills one test hole (sometimes repurposed as an outfall hole for Summit Station use) followed by up to two full-depth science holes. A planned overlap period of approximately one week between the two shifts enables on-site training, operational handover, and transfer of field knowledge between teams.

The second shift is primarily dedicated to production drilling, with the objective of completing as many science holes as possible during the remaining field season. In addition to drilling operations, the team performs routine maintenance and field repairs. The season concludes with disassembly of the drill rig and separation and packing of equipment for either transport from Summit Station or overwinter storage on site. These operations typically require approximately five days.

\section{BigRAID} 
\label{bigraid}

The BigRAID system, shown in Fig.\,\ref{fig:full_rig_cad} comprises the following systems:

\begin{figure*}[h!]
   \centering
   \includegraphics[width=1\linewidth]{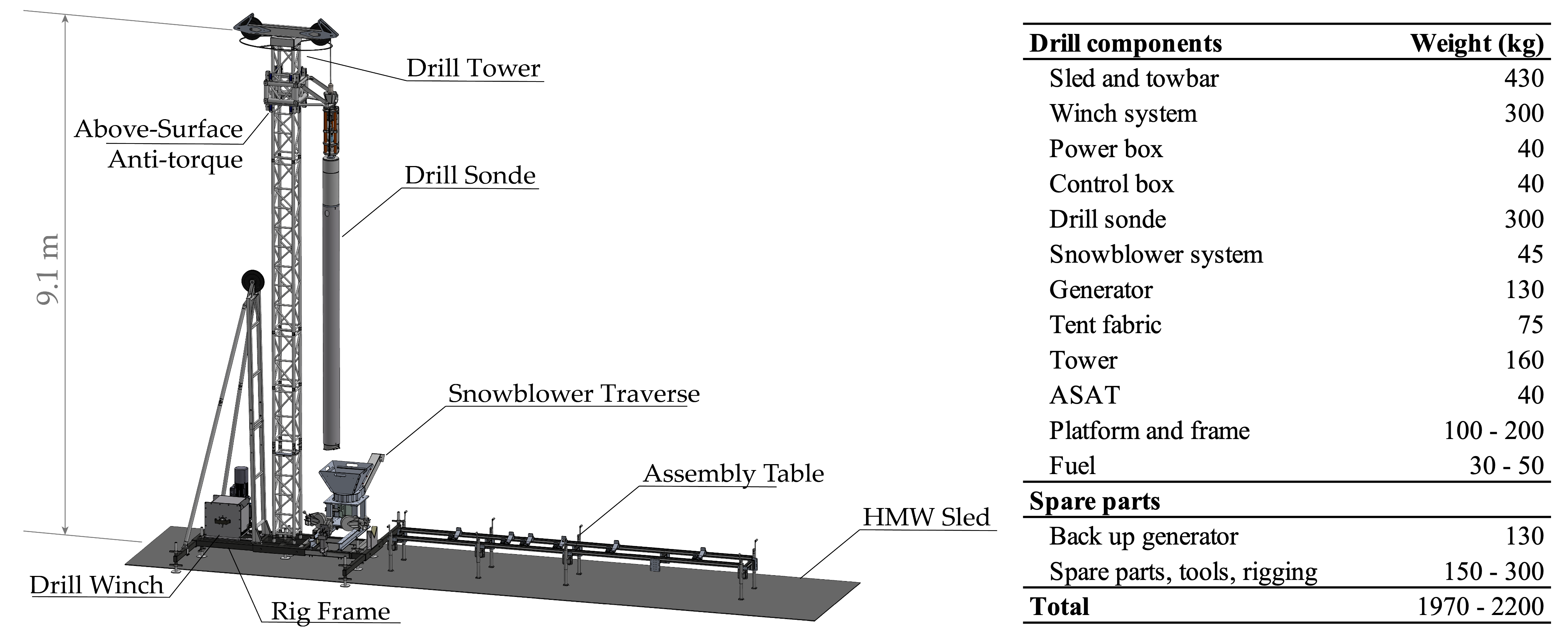}
   \caption{Left: Annotated CAD rendering of the BigRAID system. The rendering does not show the auxiliary truck winches, the power and controllers boxes, the tent frame and the tent fabric. Estimated weights of the drill system components. A range is provided for components with estimated or variable weights.} 
   \label{fig:full_rig_cad}
\end{figure*}

\begin{figure*}[h] 
   \centering
   \includegraphics[width = 1\linewidth]{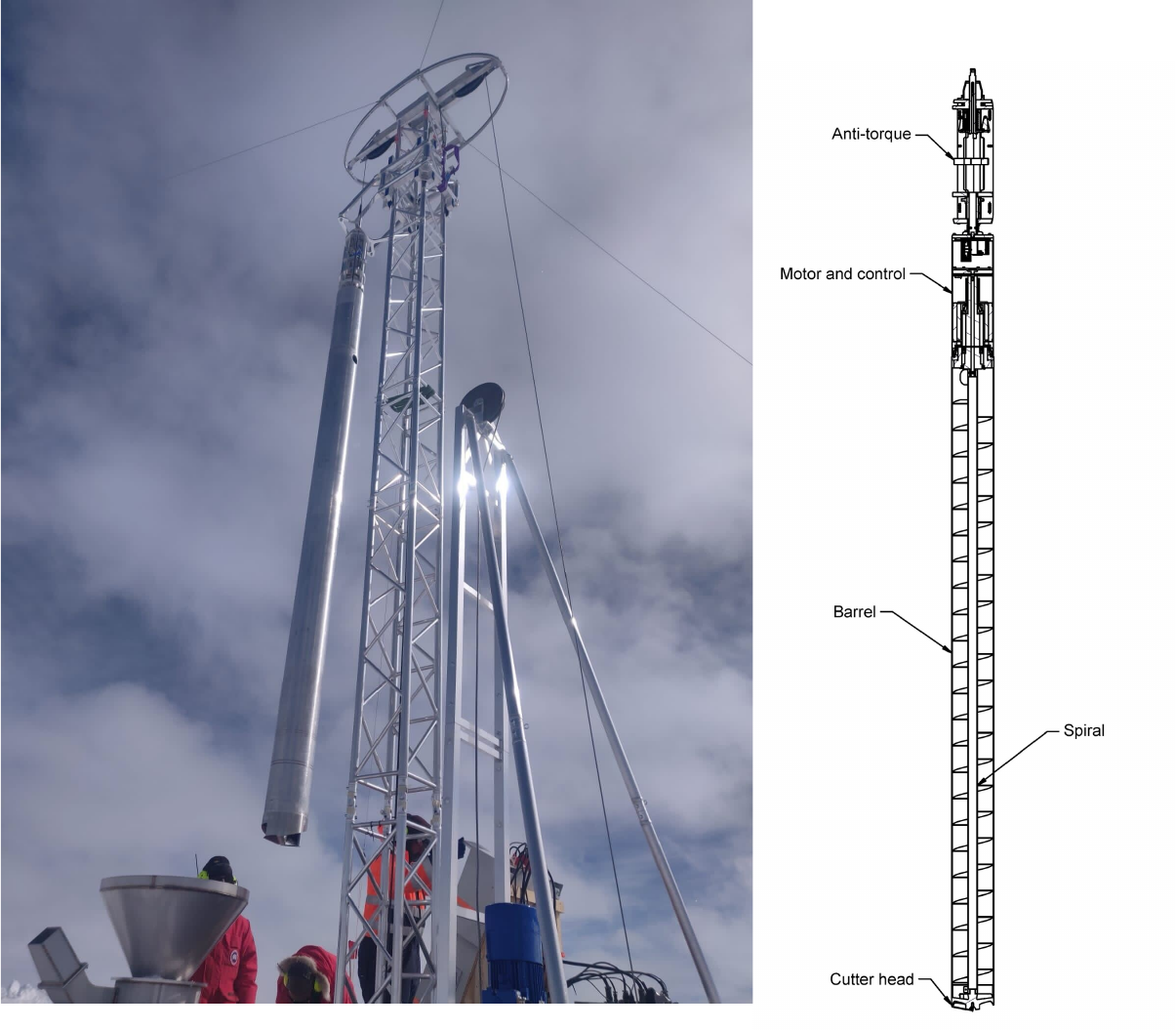}
   \caption{The drill (without the tent) and cross-sectional view of the drill sonde.} 
   \label{fig:sonde}
\end{figure*}

\begin{figure*} 
   \centering
   \includegraphics[width = 1\linewidth]{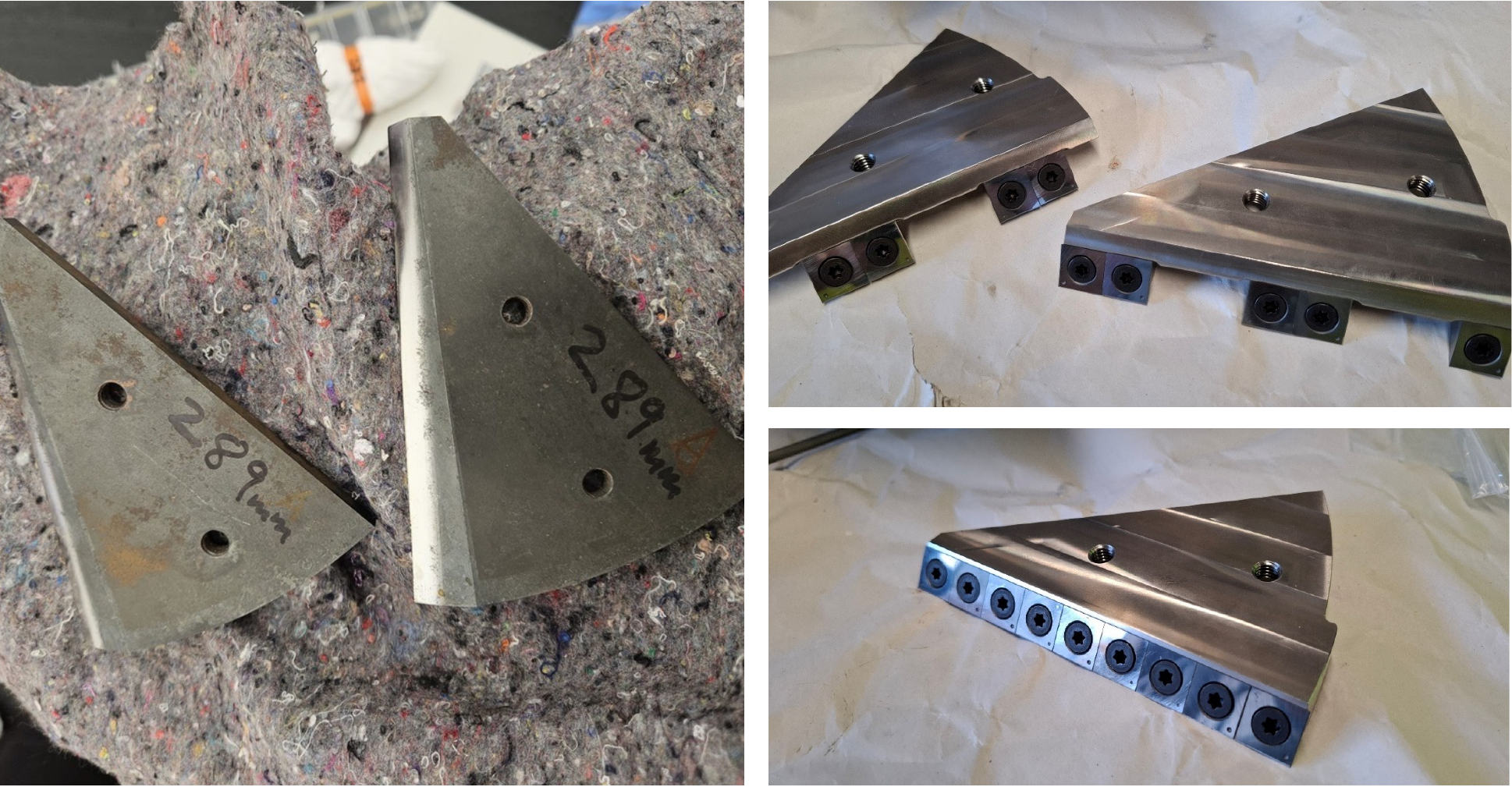}
   \caption{The three types of cutters used with BigRAID. Left: full blade cutters. Top right: staggered insert cutters. Bottom right: full-length insert cutters.} 
   \label{fig:cutters_picture}
\end{figure*}

\begin{itemize}
    \item \textbf{Drill sonde}: custom designed by the British Antarctic Survey ice drilling group with \emph{eXtreme Instrumentation}, the drill-sonde is composed of four sections, from bottom to top: the cutters, the chip-collecting barrel, the motor, and the anti-torque section (Fig.\,\ref{fig:sonde}). The brushless motor (\emph{Kollmorgen KBMS-79X05}) provides up to \SI{220}{\newton\,\meter} continuous torque and it is driven by a servo drive (\emph{Elmo Gold DC Trombone}) contained in the same housing (273\,mm diameter and 873\,mm long). The drive delivers the commanded power to the motor and communicates with the drill controls at the surface via \emph{Controller Area Network} (CAN) bus. The motor rotor is directly coupled to the 4096\,mm long barrel. A cutter head mounts to the barrel at its lower end, and can be fitted with a pair of cutters. Various configurations of cutters have been used in the past four drilling seasons: flat cutters, similar to augers used for ice fishing and manufactured from hardened American Iron and Steel Institute (AISI) A2 tool steel; full-length blades with nine carbide inserts; and staggered insert cutters, one with five carbide inserts and the other one with four. Pictures of the cutters are shown in Fig.\,\ref{fig:cutters_picture}. During drilling, the chips are pushed up the flights of a static auger that is inside the rotating barrel. The motor rotation direction and speed can be changed during while drilling, and an encoder allows the cutter to be stopped at a set position to reduce chips loss through the opening between the end of the auger and the barrel.
    
   At the top of the drill sonde, the anti-torque section (838\,mm long) provides engagement with the borehole wall using a set of retractable blades. When pushed against the ice wall, the blades counteract the torque from the driven motor and transfers torque into the cutters, which enables the sonde to drill ice.
    
    A spring-driven mechanism retracts the blades while the drill sonde is being raised or lowered in the borehole. When part of the sonde weight transfers from the drill cable to the bottom of the borehole, the spring mechanism extends the blades and push them into the hole wall. A cam plate that connects the motor and the anti-torque sections further pushes the blades against the wall with motor rotation.
    
    As the ice gets progressively harder with depth, less blade extension is needed to maintain the same amount of anti-rotation force to counteract the rotating motor. The appropriate amount of extension is needed to keep the entire sonde from spinning while allowing the sonde to freely slide downwards as ice at the bottom is removed. The blade position is manually adjusted using prescribed locating pins. This process is typically done three times per hole. A pattern of slip and stick that can be seen in the winch cable tension readout when the blades need to be retracted further.
    
    The anti-torque cannot be used in soft firn or the first 10\,m of a new hole at Summit Station. A separate Above-Surface Anti-Torque (ASAT) device described below is used for this initial section of the hole.
    
    \item \textbf{Winch system}: Custom designed by \emph{MacArtney Underwater Technology}, the winch system includes a \emph{MERMAC} winch that is rated to a Safe Working Load of 5\,kN, and a winch controller that allows the operator to preset an \emph{auto down/auto up} speed and an \emph{auto drill} speed. A 260\,m long and 6.35\,mm diameter cable (\emph{Pittsburg Wire and Cable 4-H-250}) is installed on the winch drum. The cable features a double steel armor, four 22\,AWG conductors, and has a safe working load of 10.3\,kN. Two of the conductors are used to carry power (580\,V\,DC) to the drill sonde, while the other two conductors are used for communication. 
    
    \item \textbf{Snowblower system}: The snow blowing system includes a hopper in which the snow is discharged at the end of drilling run. At the bottom of the hopper, a four-blade impeller is driven by a 3 horse power ($\approx$2.2\,kW) AC motor. The rotation of the blades both ablates the ice chips and propels the resulting fine dust away through an extended chute so that they can be dispersed outside the tent and onto the snow surface or into the air. Part of the snowblower system is also the snowblower traverse, which includes a stepper motor that positions the impeller underneath the drill sonde when the chips are ejected and moves the system away, so that it does not interfere with the drill and winch cable while the drill sonde is in the borehole.  
    
   \item \textbf{Controls}: The control box includes a programmable logic controller (PLC) that communicates with the drill sonde, and allows the operator to set rotation speed and direction of the drill. It also controls the snowblower impeller (which can be set to run at a specific rotation speed) and the traverse system movement.
   
   \item \textbf{Power box}: The power box distributes the power required to operate the drill subsystems: drill sonde, impeller, traverse.
   
   \item \textbf{Drill tower}: The drill tower (shown in Fig. \ref{fig:sonde}) stands 9.1\,m tall and it is a self-supporting structure built of welded aluminum trusses (\emph{Eurotruss ST Series}, 51\,cm, square). At the top of the tower, two sheaves guide the drill cable across the tower from the winch to the drill sonde. The tower pivots on a set of hinges at about 1.5\,m above the ground. A battery-operated \emph{Rhino model 13,500-lb} ($\approx$61\,kN) truck winch is connected to the tower above the hinge point when the system needs to be moved. Using this winch, the tower is raised for drilling; for transport, the drill sonde is laid horizontally on a sliding cart and the tower is lowered to rest on the sled without interference with the drill sonde.  
   
   \item \textbf{Above-Surface Anti-Torque (ASAT)}: A metal dolly equipped with rollers is mounted on the tower and can be positioned along its length using a second, battery-powered \emph{Rhino 13,500-lb} truck winch. Adjustable brackets on the dolly engage with a metal plate designed to capture the top of the drill sonde. The dolly is routinely used at the end of each drilling run to support the sonde while the barrel is emptied. It is also used during drilling of the upper section of the borehole, when the anti-torque system cannot yet engage with the snow and therefore cannot provide sufficient reaction torque.
   
   \item \textbf{Rig}: The complete BigRAID system is mounted on a 13.7\,m long sled fabricated from 13\,mm thick high-molecular-weight (HMW) polyethylene. Due to the extremely slippery surface of the sled material, plastic tiles manufactured by \textit{Presto Geosystems} are placed along the walking paths. The tiles are loosely connected with paracord to keep them together while allowing the sled to flex as it moves over sastrugi. The sled-based design maximizes mobility and minimizes setup time between boreholes. Once the drill system has been fully assembled at the beginning of the season, only a few operations are required to reconfigure it between drilling and transport setups. This allows the entire rig to be rapidly (within 3--4\,h) relocated between holes by towing it with a small tracked vehicle, such as a \emph{Pisten Bully 100}. The rear section of the sled houses the primary power systems, including a \emph{Europower EP18000TE} generator and a backup generator, several storage boxes containing tools, spare parts, and field maintenance equipment. The front section supports the main structural frame of the drill system, including the operator platform, the control and power electronics enclosure, the snowblower system, and the drill tower assembly. The total weight of the rig is estimated to be around 2000 kg as illustrated in Fig. \ref{fig:full_rig_cad}.

    \end{itemize}

\section{BigRAID Operation}
The BigRAID drill rig is self-contained, and can be moved on a sled to the target location. Once in position, some preparation steps are required before drilling operations start. The preparation steps include leveling the rig using screw outriggers, raising the drill tower and tent, attaching snow anchors, hanging the drill from the tower, and performing a test of all drill subsystems. 

In our specific context, the first 10\,m of the hole is referred to as the \emph{shallow section}, and the remaining deeper part is the \emph{deep section}. We will describe drilling in the deep section first, as drilling the shallow sections follows essentially the same procedure with some additional steps.

\begin{figure*}[h]
    \centering
    \includegraphics[width=1\linewidth]{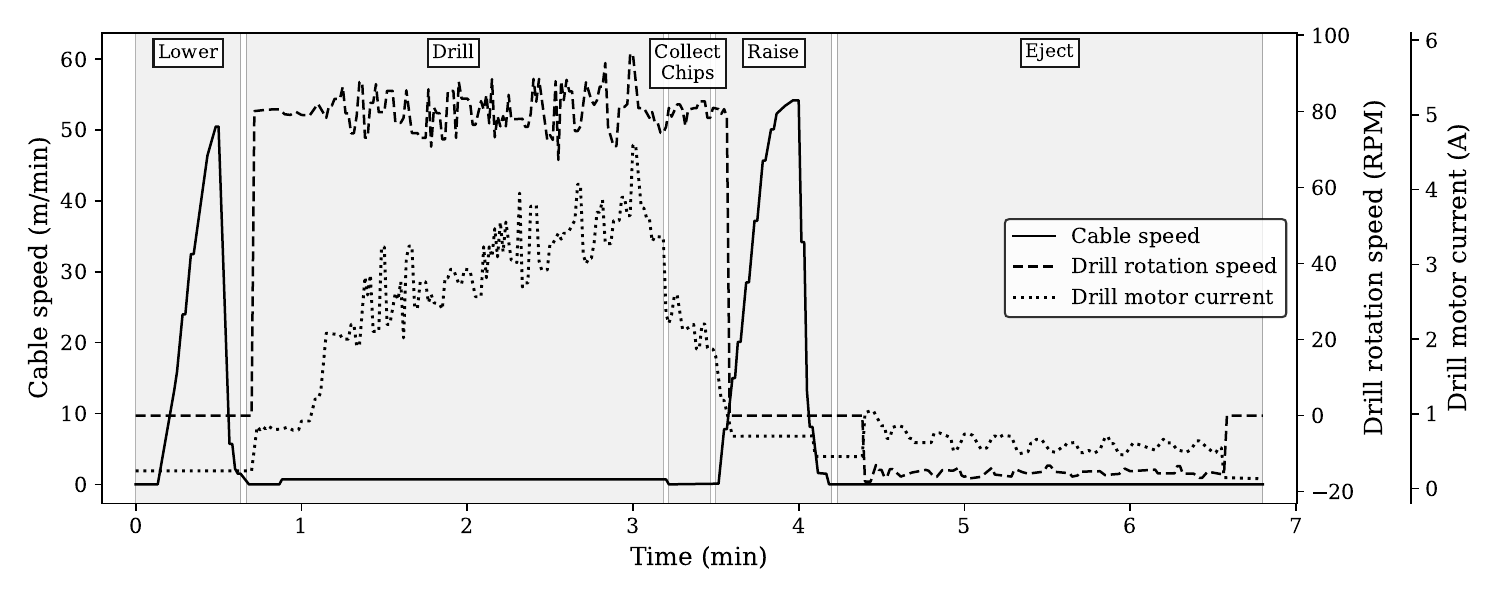}
    \caption{Drill and winch parameters for a typical drill run from 16 to 17.8\,m (1.8\,m cut) during the 2025 field season. This run was performed using the automation function.}
    \label{fig:run_plot} \label{drill-seq}
\end{figure*}

The BigRAID software records operational parameters such as the winch load and speed, and the drill motor current while it is active. These can be viewed in real time to make decisions in the field, and are also recorded for later analysis. Fig.~\ref{fig:run_plot} shows the winch cable speed, drill rotation speed, and drill motor current over time for a typical drill run in the deep section.

Initially, the drill sonde is hanging empty over the hole, ready for a drill run to start. It is then lowered to the bottom of the hole. A function on the winch controller allows the drill to automatically lower to the last \emph{lock depth} value, which is set at the end of drilling in the previous run. A small offset, typically configured to around 20\,cm, is subtracted from this depth to account for any ice chips that may have fallen out of the barrel while raising the drill sonde.

Drill rotation is then started, and the winch is lowered at a constant slow speed to achieve the intended \emph{rate of penetration} (ROP). The drill motor current, which is roughly proportional to the torque, clearly shows the point at which the drill has been lowered past the offset and is now actively drilling ice. At this point, the anti-torque system, which grips the hole to counter the rotational force created by the drill head, fully engages through a mechanical process triggered through the drill head rotation and the load coming off the winch cable. Drilling continues until either a desired cut length has been reached, or until the motor current exceeds the safe limit of the drill cable at around 5\,A. 

After the drilling phase, winch movement is stopped, and the drill is left to linger at the current depth, while rotating, to collect remaining chips at the bottom of the hole into the barrel. The current depth is locked on the winch so the drill can be lowered to this position in the next run. The drill is then raised back to the surface. To reduce the likelihood of the drill getting stuck at the bottom of the hole, drill rotation is only stopped a short time after the winch starts moving the drill upwards. This ensures that the drill is never completely stationary while in the hole.

At the surface, the drill engages with the ASAT system on the tower, and the snowblower impeller is started and it is brought into position below the drill by the traverse system. The drill then starts spinning counter-clockwise (opposite the drilling direction). This allows the chips to fall from the barrel into the snowblower and be ejected. Once the barrel is empty and the snowblower has moved back to its parking position, the drill is ready for the next run.

When drilling the initial shallow section, the drill cannot yet engage the hole with the anti-torque system, either because it is still above the hole or because the initial snow and firn layers are too soft. Therefore, operators attach extensions to the drill stack that couple the anti-torque to the ASAT unit on the tower.

The procedure above can be executed by the operators directly using the electronic controls to set the winch speed, drill rotation speed etc.  For the deep region, these stages can also be executed automatically by the system. The automation function is described in more detail in the BIGRAID UPGRADES section.



\section{Drill performance}
\begin{figure*}[!htb]
\centering{\includegraphics[width=1\linewidth]{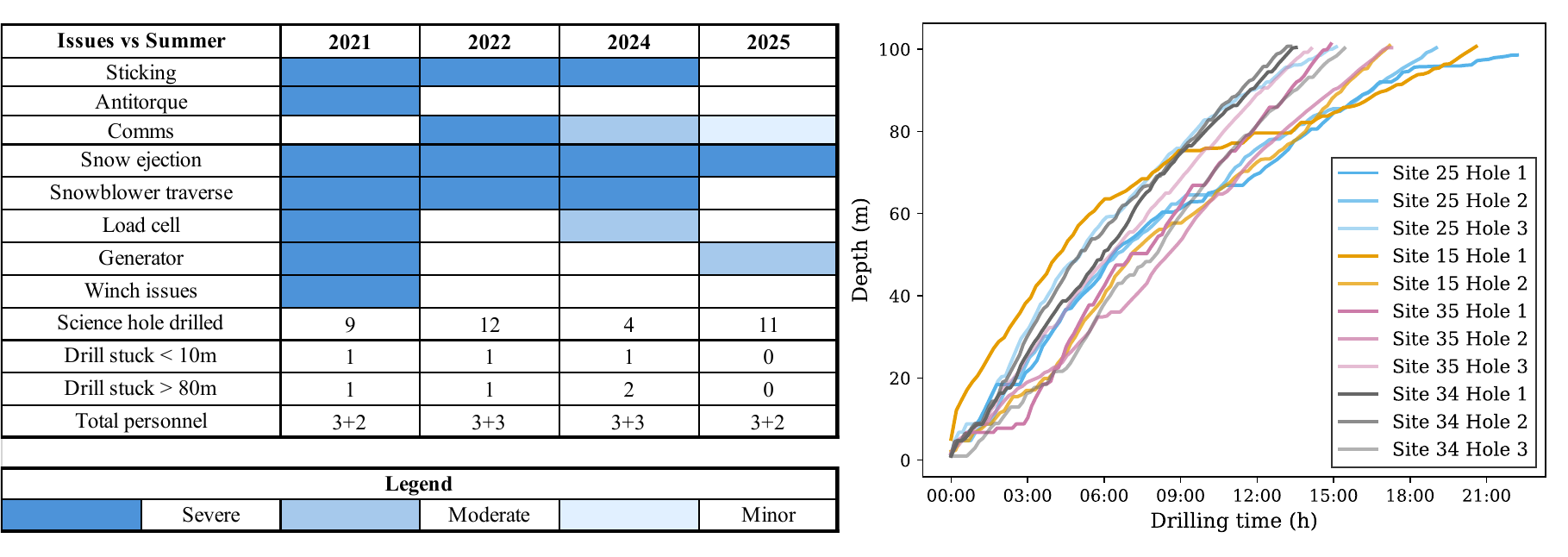}}
\caption{Left: issues encountered in four seasons of drilling. Color intensity represents impact of issue on drilling operations. Right: depth vs time for the eleven science holes accomplished during 2025. During smooth operations, drilling a borehole requires about 40 liters of fuel.}
\label{fig:performance}
\end{figure*}

Fig. \ref{fig:performance} summarizes issues and performance of the BigRAID drill over four drilling seasons and shows the performance of the drill for the eleven holes drilled in 2025. 

Due to delays related to the COVID-19 pandemic, the drill was assembled in the field for the first time in late May 2021. Despite the system still lacking several essential components, including a load cell, the team successfully completed the third science hole in one single day. However, at 99\,m depth, the anti-torque system slipped, damaging the cable and requiring it to be cut and re-terminated. 

During the same season, the drill became stuck just below the surface. The sonde was partially excavated manually and eventually recovered using heavy equipment fitted with a jib boom. This sticking event was attributed to above-freezing air temperatures, which caused the drill sonde metal surfaces become wet and refreeze to the borehole. Warm conditions also led to the formation of ice blocks inside the auger flights, periodically preventing operation until the ice could be removed mechanically or thermally. In addition, the original snowblower traverse system proved too slow and unreliable for field operations and was reverted to manual operation. Despite these significant challenges, nine science holes were completed during this first season with a minimal crew, a remarkable outcome for a drill system still in its infancy.

To mitigate sticking issues observed in 2021, during the 2022 season we introduced a white tent enclosure to reduce solar heating of the drill structure. Although the tent improved working conditions and reduced direct heating of metal surfaces, sticking events still occurred twice: once at approximately 10\,m and once near 92\,m depth. In the shallow event, the drill was again excavated manually and extracted using heavy equipment. In the deeper event, glycol was introduced into the borehole to release the drill. Because little was known about the volume required to free such a large sonde, and with limited time remaining before season end, an entire drum of heated \emph{DOWFROST Propylene Glycol HD} (the only glycol available at Summit Station at the time) was delivered into the borehole through a makeshift hose system. The drill was successfully recovered within a few hours. Another major operational issue during the 2022 season involved unreliable communication between the main PLC and other subsystems. Frequent communication faults required repeated restarts of the controls throughout the day, resulting in substantial operational downtime. Nevertheless, the team completed twelve science holes that season. While the tent substantially improved operator comfort and environmental protection, it also complicated inter-hole operations: moving the system required three people and imposed additional wind-speed limitations for safe raising and lowering of the tent structure.

In 2023, the RNO-G collaboration elected to suspend drilling operations in order to focus on system development and reliability improvements. Significant progress was made in both controls and mechanical systems. The control software was extensively revised, substantially reducing communication errors between subsystems. Automated drilling (\emph{auto mode}) was further developed and integrated into regular operations, while data logging capabilities were expanded considerably. An orientation sensor was also added to the motor housing to monitor sonde rotation and provide warning of excessive twisting that could potentially damage the cable. Several mechanical upgrades were implemented on the sonde itself. The cutter head was redesigned to accommodate both traditional flat cutters and newer insert cutters, and sharp external edges were tapered to reduce snagging and improve handling. As part of these modifications, the chamfer geometry at the cutter head/barrel interface was also changed. In addition, a dedicated field-deployable glycol delivery system was designed and prepared in anticipation of future sticking events.

\begin{figure*}[!htb]
\centering{\includegraphics[width=1\linewidth]{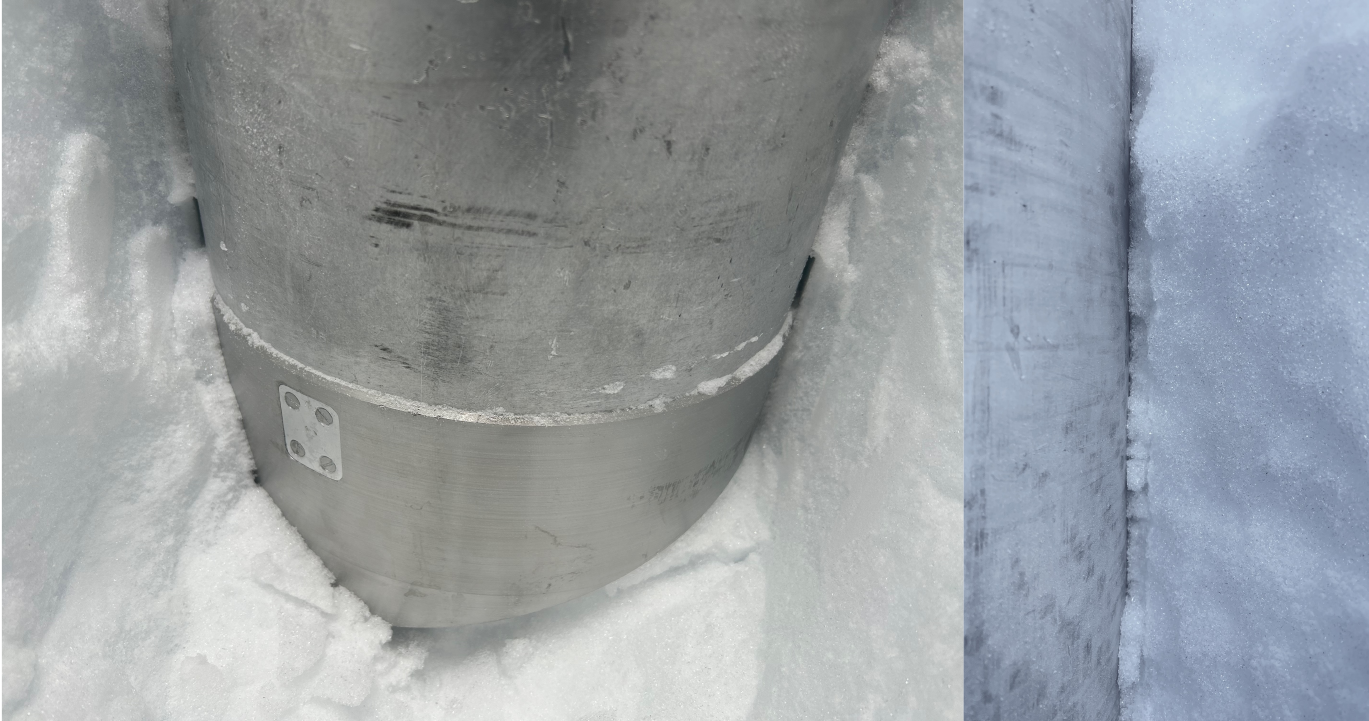}}
\caption{Images of the stuck drill sonde, partially dug out (left), and barrel-ice interface (right).  Chips which are not collected inside the barrel are stuck between the drill sonde and the ice wall, jamming the cutter head due to the protruding edge.}
\label{fig:stuck_drill_head}
\end{figure*}

\begin{figure*}[!htb]
\centering{\includegraphics[width = 1\linewidth]{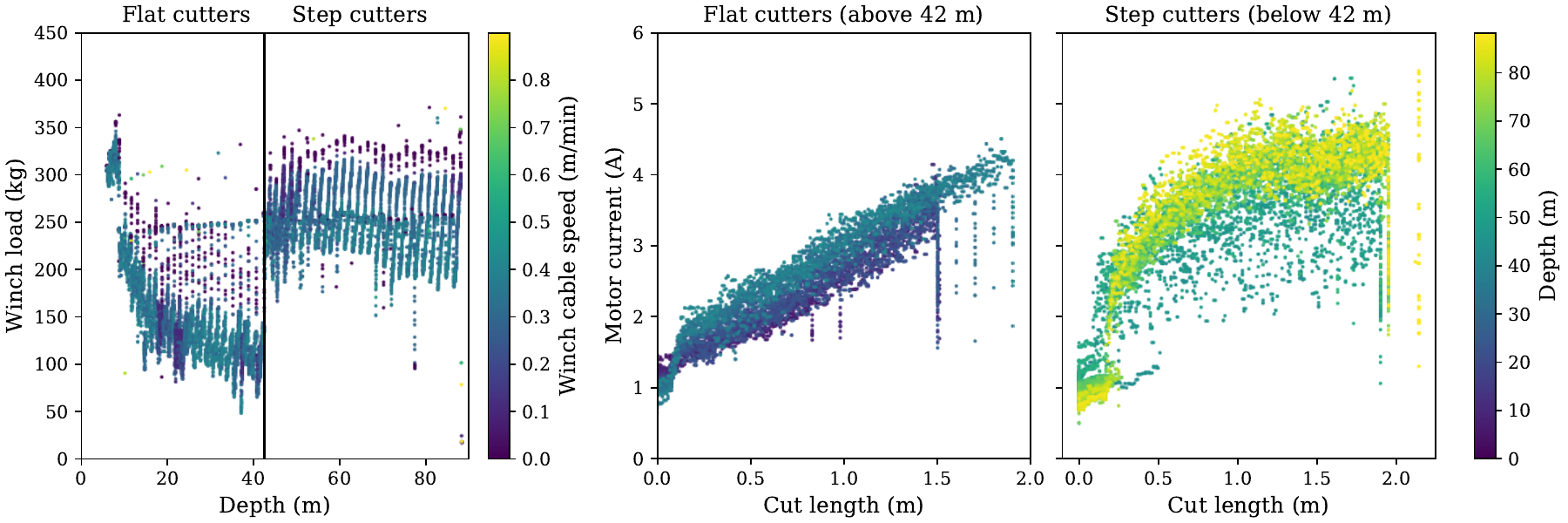}}
\caption{Comparison of performance of the drill with flat and staggered insert cutters. The leftmost panel shows the winch load (as measured by the winch load cell) vs depth. The center and right panels show the motor current pattern as a function of cutting progress within the run and depth.}
\label{fig:cutters_performance}
\end{figure*}

\begin{figure*}[!htb]
\centering{\includegraphics[width = 1\linewidth]{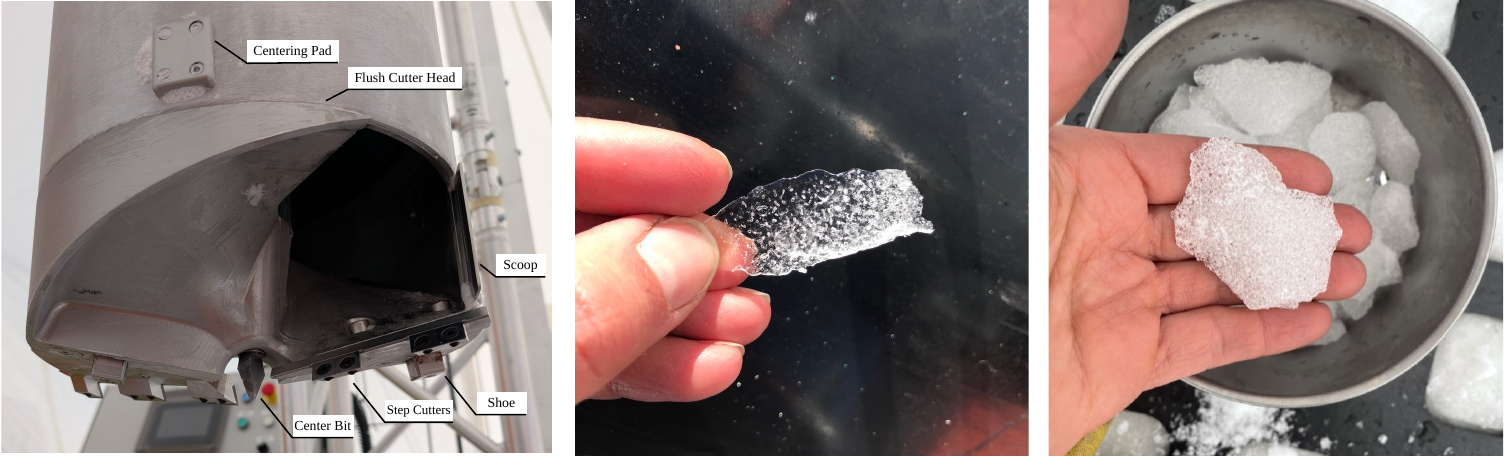}}
\caption{Left to right: New cutter head with staggered insert cutters; chips produced with flat cutters; chips produced with staggered insert cutters}
\label{fig:head_chips}
\end{figure*}

Drilling operations resumed in 2024 with the upgraded system. Automation improvements allowed the \emph{auto mode} to be adopted operationally from the beginning of the season. However, sticking problems became even more severe, with the drill becoming stuck three times within a period of approximately three weeks, including both shallow and deep events. In the deep sticking events, 50 to 60 liters of heated 99\% propylene glycol were delivered efficiently down-hole using the newly developed glycol deployment system, successfully freeing the drill each time. When the drill became stuck at approximately 6\,m depth, the team manually excavated the drill sonde (see Fig.~\ref{fig:stuck_drill_head}). Direct inspection of the head strongly suggested that the cutter-head \emph{rim} was the primary cause of sticking. Ice chips not captured and transported into the barrel appeared to accumulate around this rim and jam the sonde in the borehole, an effect likely exacerbated by the modified chamfer geometry introduced in 2023. The chamfer orientation was subsequently reversed, and the following science hole was completed successfully in record time. This hole also marked the first successful deployment of the new staggered cutters which produced much larger chips as shown in Fig.\,\ref{fig:head_chips}. Fig.\,\ref{fig:cutters_performance} shows the change in drilling parameters when the cutters were swapped at 42\,m depth. A higher load on winch (lower weight on bit) and higher, more erratic motor currents were recorded. In spite of the less predictable and higher motor currents, cut lengths of up to 1.8\,m per run were achieved at 80\,m depth. This mechanism is not fully understood.


Based on these findings, a new cutter head design was manufactured for the 2025 season with an external profile flush with the barrel diameter. This modification proved highly successful: for the first time since deployment of the drill, the season was completed with zero sticking events, even under stress-test conditions such as loss of power while actively cutting, and overfilling the barrel. Eleven full-depth science holes were completed during the season, as shown in Fig.\,\ref{fig:performance}. 

The most impactful upgrades implemented between 2024 and 2025 are described in the following section.

\section{BigRAID Upgrades}
\label{sect:upgrades}

\subsection{Automation}

The possibility to fully automate the drilling process has been one of the reasons BigRAID was chosen as a drill, as RNO-G is considered the pathfinder to a much larger installation of antennas part of the IceCube-Gen2 Neutrino Observatory \citep{Aartsen_2021}. To build this future array in a reasonable time frame, multiple drills will need to be operated simultaneously with a team as small as possible. The automation function of the BigRAID, developed and used in the field starting 2024, can be used to largely automate the manual drilling procedure. In this mode, the drillers define the operational parameters of drilling and supervise the drill. The system can only be used after the initial shallow section of the hole has been drilled, as the initial section requires manual steps to engage the ASAT.

\begin{figure*}
    \centering
    \includegraphics[width=1\linewidth]{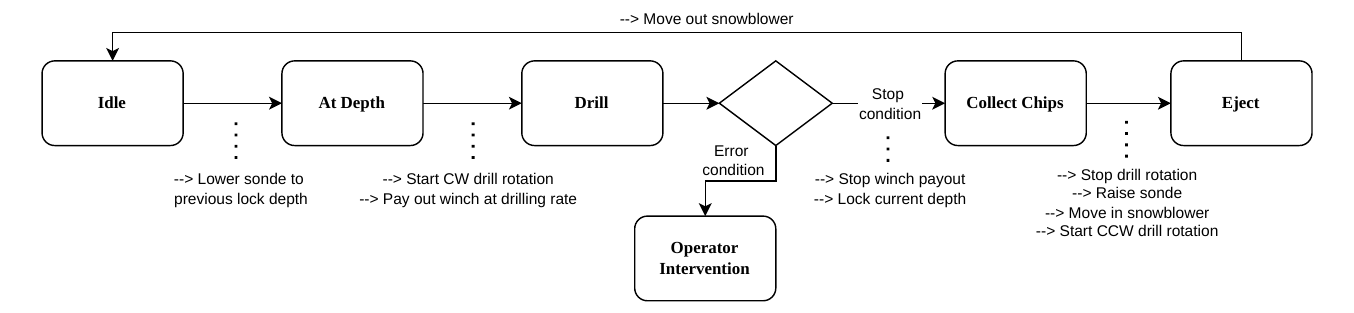}
    \caption{Simplified BigRAID automation state machine.}
    \label{fig:automation_flowchart}
\end{figure*}

Figure\,\ref{fig:automation_flowchart} shows a simplified state machine that is traversed repeatedly for every drill run. This system generally requires operator intervention in case of any error condition. These include unexpected component state, such as the drill not maintaining the requested rotation speed, exceeding system limits such as maximum drill motor current, and some monitoring conditions such as slippage of the anti-torque system as measured by an \emph{Inertial Measurement Unit} (IMU) included inside the motor housing.

In the idle state of the system, the drill sonde hangs above the hole fully ejected. The system then lowers it until it is at the last drilled depth with a small offset to account for potential dropped ice chips from the previous run. Next, drill rotation is started, and the system pays out the winch at the configured drilling feed rate.

Multiple different stop conditions are used to stop the drilling phase of a run. The system was originally designed to drill a specific \emph{cut depth}. However, the maximum cut depth that would fill the barrel cannot be reached under typical conditions. This is because the torque, and therefore current, required to drill increases as the barrel is filled. The characteristics of the drill cable limit us to approximately 5\,A of continuous current. Automation uses this \emph{motor current limit} as an additional stopping condition, with the goal of maximizing the cut depth of each run. 

The automation also estimates the \emph{weight on bit} during drilling based on the known drill weight, the load measured by the winch and an estimated weight of the ice already drilled in this run. The drill phase is also stopped if this value exceeds a configured limit in order to prevent unspooling drill cable into the hole if the drill is not making sufficient progress.

\subsection{Snowblower traverse system}
\begin{figure*}[!htb]
    \centering
    \includegraphics[width=1\linewidth]{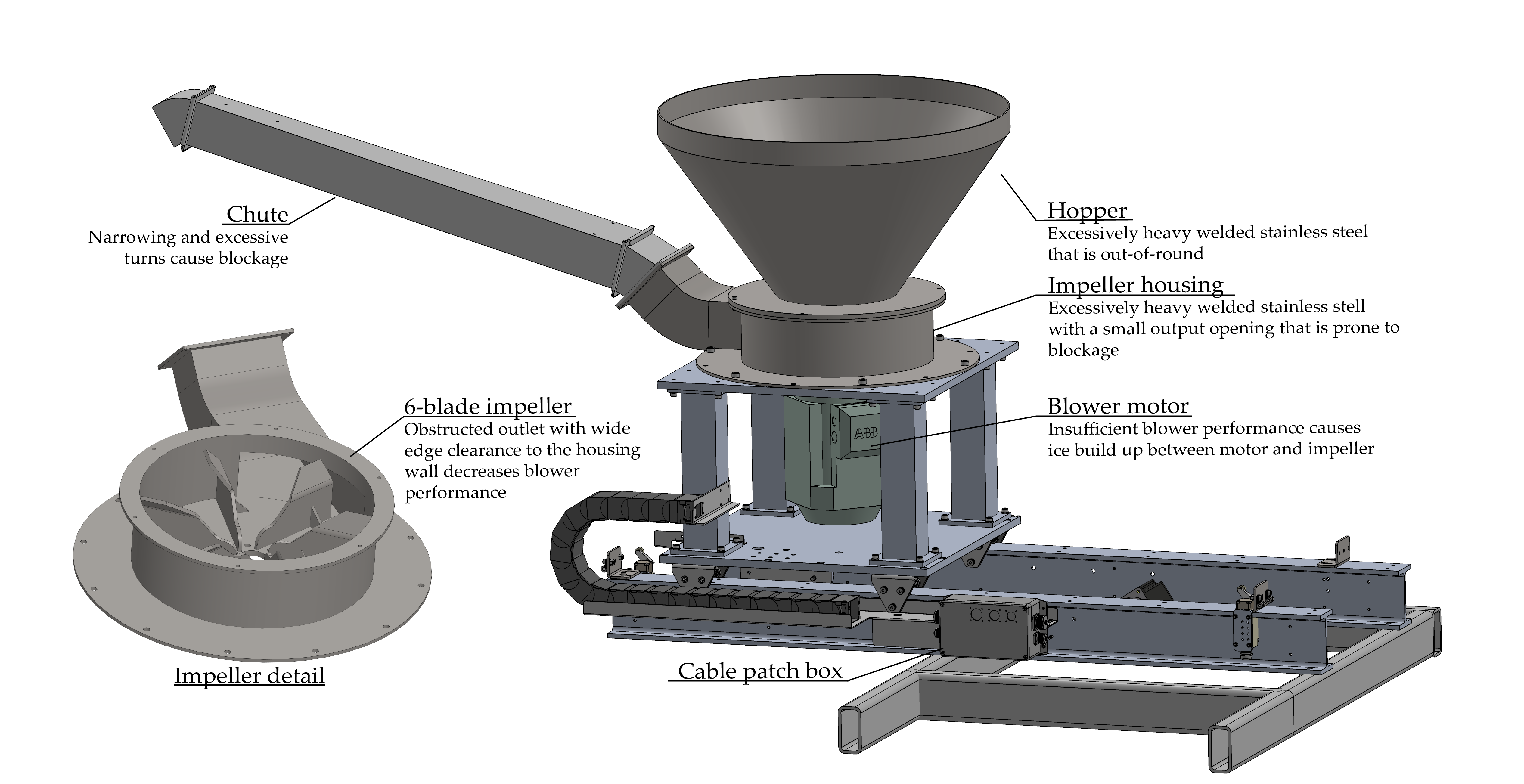}
    \caption{CAD rendering of the redesigned snowblower traverse system used in the 2025 field season.}
    \label{fig:snowblower_2025}
\end{figure*}

The snowblower traverse system, as shown in Fig.\,\ref{fig:snowblower_2025}, is used to discard the ejected snow and ice chips from the drill. Once the drill has resurfaced from the hole after each drilling run, the motor is rotated at typically 20 revolutions per minute (rpm) counterclockwise (opposite to the drilling direction) to expel the collected snow and ice chips into the snowblower that is now positioned underneath the drill. The snowblower is equipped with a rotating impeller that blows the collected ice chips away from the drill rig. At the end of this process, the snowblower is moved, or traversed, aside to make way for the drill to descend into the hole for the next drill run.

The traversing mechanism was originally designed using a stepper motor-driven lead screw rail and a supported linear rail that were set up in parallel. These 1.6 meter-long precision rails were placed approximately 0.5 m apart and required sub-millimeter parallel alignment to work without binding. This alignment could not be maintained as the underlying structure for the rails twisted and bent as the drill rig was moved across the glacier, leading to binding of the traverse system. The lead screw was also undersized for the long span and load such that visible buckling was observed during use. The buckling was more pronounced as the rails tended to bind and further increased the compressive forces in the lead screw.

These issues were addressed in two ways. First, the lead screw was replaced by the Avid CNC PRO Rack and Pinion Drive 2.0. The rack gear was attached to the stationary support structure, and the pinion gear was attached to the moving snowblower. The die spring that provides the meshing force between the pinion and rack gears compensates for slight bending misalignment of the support structure. This drive system also includes a belt-driven 3.2:1 gear reduction to allow usage of smaller stepper motors, as needed, at the cost of slightly slower traverse speeds but higher available torque. The gear reduction also serves as a braking system that keeps the snowblower from rolling due to uneven terrain. Since the original drive components were intended for indoor use, some fasteners and the drive belt were replaced with environmentally-compatible counterparts.

Second, the precision linear rails were replaced by rollers similar to the ones on a gantry hoist trolley. Sealed ball bearings mounted on custom brackets served as rollers while acetal pads served as side-loading wear surfaces that also constrained the snowblower on the supporting structure while allowing sufficient clearance for slight deformations in the structure. Nitto Permacel 212 glass cloth tape was readied in case the rollers loss traction on the supporting structures. This was not observed through the drilling season, and even wear around the rollers supports this observation. Through these design improvements, the snowblower traverse system functioned smoothly and reliably, and required little maintenance throughout the season. Additionally, the total travel time of the traverse during a drill run was reduced from approximately 120\,s to 20\,s.

\subsection{Drill sonde modifications}
The mechanical and electrical interface between the electromechanical winch cable and the drill sonde is the cable head, which originally, was the \emph{TrustLink Stress Terminator} (Type I-S) by \emph{MacArtney Underwater Technology}. The installation of this cable head requires epoxy-potting the winch cable into the termination, which is not field-serviceable. In the case of winch cable damage or failure, operators will have no practical means of repairing the cable termination in the field. The new field-serviceable \emph{EverGrip} cable head, model 117-2C-L, made by \emph{PMI Industries}, replaces the original termination.

A two-pin clevis is incorporated to allow the electrical cable that passes through the \emph{EverGrip} cable head to bend freely as the clevis is rotated. The clevis rotates every time the drill sonde is lifted or lowered between its vertical and horizontal configuration. The sonde is vertical when operational and horizontal during transport and servicing. 

A mechanical and an electrical slip ring were also incorporated into this design update to further increase the drill robustness. The mechanical slip ring incorporates a tapered bearing pair to withstand the sonde weight at a safety factor of 5. The flange-mount electrical slip ring is made by \emph{IEC Corporation} (FMS series). With these two slip rings, the sonde can rotate freely with respect to the winch cable. This is a critical feature that prevents severe damage to the winch cable whenever the anti-torque system fails while the drill motor is rotating. In such a scenario, the multi-strand winch cable may unravel, or the electrical cables may twist beyond their stress limits, both of which leads to catastrophic failures to the sonde. 

\subsection{Orientation and tilt sensor}
\begin{figure*}[!htb]
\centering{\includegraphics[width = 1\linewidth]{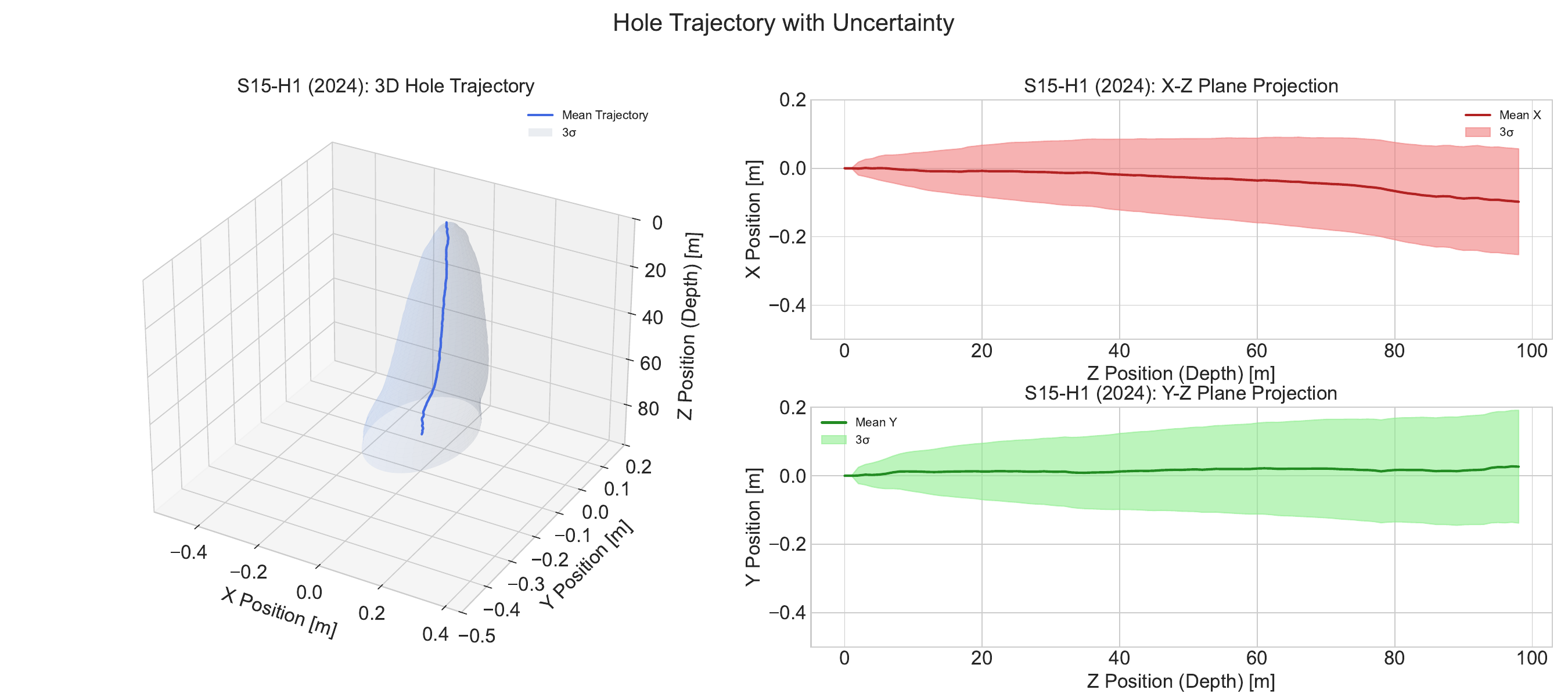}}
\caption{Hole inclination vs depth Station 15 Hole 1 (2024). The trajectory is reconstructed from angles measured by the IMU while the drill is being lifted to the surface. Left: The 3D reconstructed trajectory. Top Right: The deviation in \emph{X} direction along the depth of the hole. Bottom Right: The deviation in \emph{Y} direction.}
\label{fig:S15-H1_inclination}
\end{figure*}

\begin{figure*}[!htb]
\centering{\includegraphics[width = 1\linewidth]{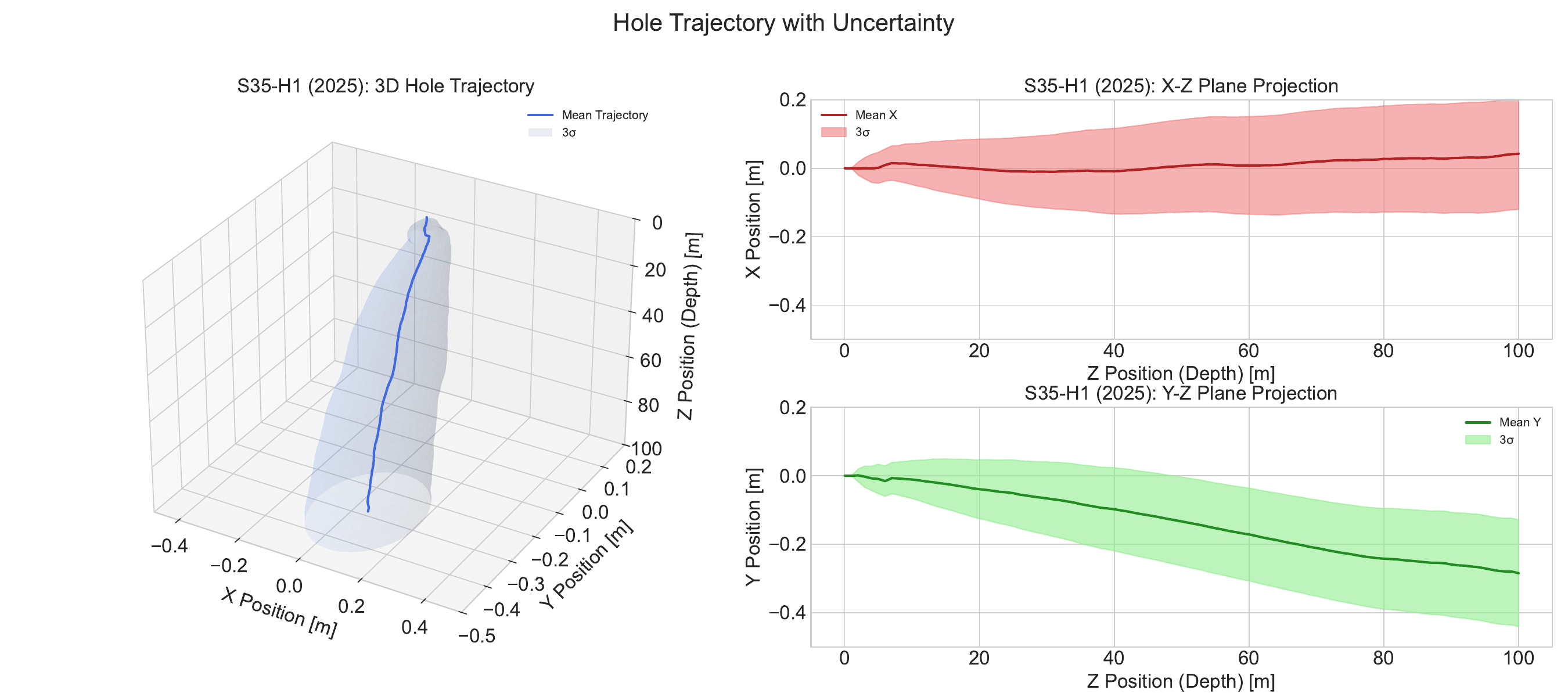}}
\caption{Hole inclination vs depth Station 35 Hole 1 (2025). The trajectory is reconstructed from angles measured by the IMU while the drill is being lifted to the surface. Left: The 3D reconstructed trajectory. Top Right: The deviation in \emph{X} direction along the depth of the hole. Bottom Right: The deviation in \emph{Y} direction.}
\label{fig:S35-H1_inclination}
\end{figure*}

In 2023 we added a ML5-AR \emph{Inertial Movement Unit} (IMU)/tilt sensor to the motor section of the drill sonde. The sensor features a triaxial accelerometer and a triaxial gyroscope and provides a measurement of pitch, roll, angular rate and acceleration. The main motivation behind this upgrade was to allow the operator to monitor for excessive anti-torque slippage, which, due to the lack of slip ring in the system at the time, could cause damage to the cable. The data from the IMU is displayed on the drill controls and is logged with the other data and allows to determine the borehole geometry. From pitch and roll we can reconstruct the \emph{X} and \emph{Y} coordinates of the hole center and reconstruct the hole trajectory. Data can be analyzed in subsets (up-going, down-going, cutting, not cutting). While the holes are generally quite straight (with a deviation of the center by less than one third of the diameter at full depth - see Fig. \ref{fig:S15-H1_inclination}), some holes exhibit a larger deviation (see  Fig. \ref{fig:S35-H1_inclination}). Understanding the correlation between inclination and operational drill parameters will be part of future studies. 

\subsection{Tent lift kit}
Before moving to the next hole, the drill tower must be lowered into its transport position. After the addition of the tent enclosure, this operation became significantly more labor-intensive, requiring a minimum crew of three people. Two operators were needed to manually lift the front section of the tent frame until it could follow the motion of the tower, while a third operator controlled the truck winch used to lower the structure. Because the long-term operational goal for BigRAID is a two-person drilling crew, this requirement imposed a substantial logistical burden on inter-hole operations.

For the 2025 season, the tent deployment system was substantially improved through the addition of a lift kit consisting of a 4:1 pulley system combined with an ascender. One end of the system is attached to the oval ring located at the top of the tower, while the other is connected to the front section of the rig frame. This configuration allows a single operator to control the motion of the tent frame by tensioning the pull rope and raising fully the tent frame against the tower. The hinge-locking pins securing the tower can then be safely removed and finally the tower and tent assembly can be lowered in a controlled manner. This relatively simple mechanical addition significantly streamlined inter-hole operations and successfully reduced the required personnel from three operators to two.

\begin{figure*}[!htb]
\centering{\includegraphics[width = 1\linewidth]{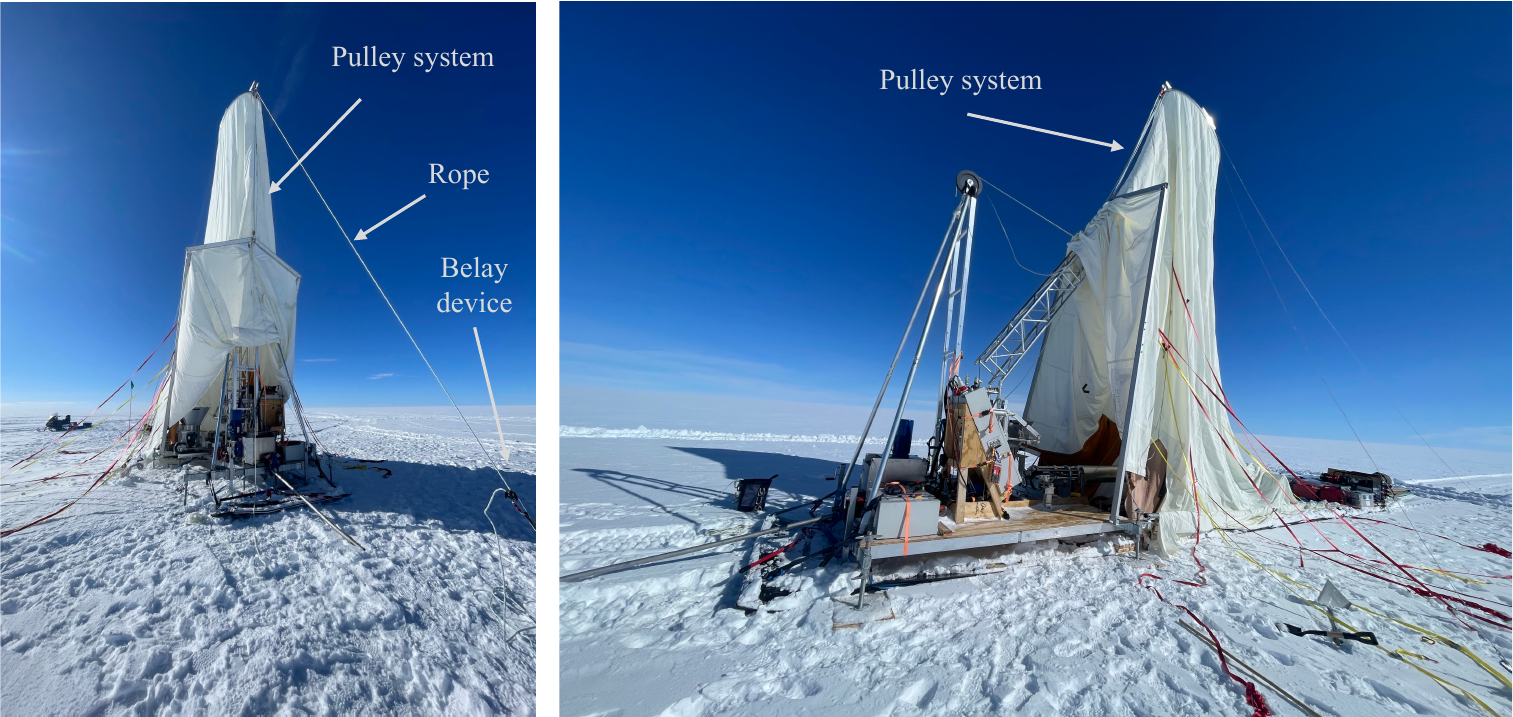}}
\caption{Preparing the tent for moving. Left: the tent is lifted using a 4:1 pulley system connected between the top of the tower and the tent frame front. An auto-locking belay device on the pull side of the line and anchored in the snow allows for the operator to step away as needed. Right: the tower is lowered using the truck winch. Tent fabric and front frame rests on the tower.}
\label{fig:tent_ops}
\end{figure*}

\subsection{Drill recovery kit}

Since the drill got stuck several times during its development years, we have developed an emergency recovery kit that can be deployed quickly in case of need. The kit includes:

\begin{itemize}
    \item \textbf{95\% Propylene Glycol}: a 55\,US\,gal ($\approx$208\,L) drum of 95\% propylene glycol, kept outdoors at Summit Station.
    \item \textbf{Glycol Corrosion Inhibitor}: 3\,US\,gal ($\approx$11\,L) of glycol corrosion inhibitor with a colorant to dye the glycol and prevent damage to drill components.
     \item \textbf{Jerry cans}: four 20\,L jerry cans are allocated to transfer the max amount we foresee to be used in a sticking event based on our past experience.
    \item \textbf{Transfer pump}: a hand-operated pump to move glycol from the barrel (which is stored outside) to the jerry cans.

    \item \textbf{Heating blankets}: four 80\,W heating pads to warm up the jerry cans either indoor or with a small portable generator.
    \item \textbf{Hose}: sections of 25\,mm \emph{Forest Lite} hose from \emph{Kuriyama} with \emph{National Pipe Straight Hose} (NPSH) fittings, including four 30\,m, one 15\,m, one 7.5\,m, and one 3.3\,m sections. The hose sections can be joined together to reach a wide range of the borehole depths.
    \item \textbf{Mechanical hose holder}: a custom holder to keep the hose in place when deployed and to squeeze out the glycol when reeling in the hose from the borehole.
    \item \textbf{Sprayer}: a custom shower-head like attachment that is connected to the bottom of the hose to more precisely spray the glycol toward the borehole wall. This allows for a more efficient delivery of glycol compared to a concentrated stream directly out the hose.
\end{itemize}

In addition to the glycol delivery kit described above, a suite of wireline recovery tools has been acquired to allow retrieval of the drill sonde in the event of a cable failure, either at the cable head or at another location along the cable. The top of the drill sonde incorporates a fishing-neck profile compatible with commercially available downhole pulling tools developed for the oil and gas industry. This allows the sonde to be recovered using standard wireline fishing techniques and reduces the need for specialized recovery equipment. The current recovery tool set includes:

\begin{itemize}
\item \textbf{Cable grips:} Designed to be wrapped around the cable to secure and support the drill sonde if the winch becomes inoperable or the cable is damaged above the grip location.

\item \textbf{Bowen-type rope socket:} Provides a secure mechanical connection between the wireline and a downhole tool. Depending on the failure scenario, it can be connected to a damaged cable and the GS pulling tool to recover the drill sonde, or to the mechanical Spang jar and cable grab for recovery of loose cable within the borehole.

\item \textbf{GS pulling tool:} Designed to engage and retrieve downhole equipment equipped with a compatible fish neck profile. The tool provides the primary means of mechanically engaging the drill sonde if the cable connection is lost.

\item \textbf{Mechanical Spang jar:} Provides controlled impact forces in both the upward and downward directions to assist in freeing or engaging downhole equipment. It consists of two interlocking sliding sections that operate similarly to a slide hammer, generating an impact when the tool reaches the end of its stroke.

\item \textbf{3-leg wireline grab and spear:} Designed to recover broken or lost wireline from the borehole by engaging and retaining loose or coiled cable. The tool can be used in combination with the Spang jar to assist in securing cable that cannot otherwise be retrieved.
\end{itemize}

The cable grips are manufactured by \textit{PMI}, while the remaining recovery tools are manufactured by \textit{Brace Tool}.

\subsection{Remaining issues in drilling}

Despite the major improvements achieved in 2025, some temperature-related operational challenges remained. The snowblower system continued to be prone to icing and snow accumulation. Application of \emph{Nanomyte Super Ai} coatings successfully mitigated ice buildup on internal surfaces, although snow could still clog the chute under certain conditions. In addition, one event involving the formation of large ice blocks inside the barrel was observed. The most likely explanation is that residual snow remaining inside the barrel after drilling partially melted and subsequently refroze during transport between holes while the sonde remained horizontal for extended periods.

\begin{figure*}[!htb]
\centering{\includegraphics[width = 1\linewidth]{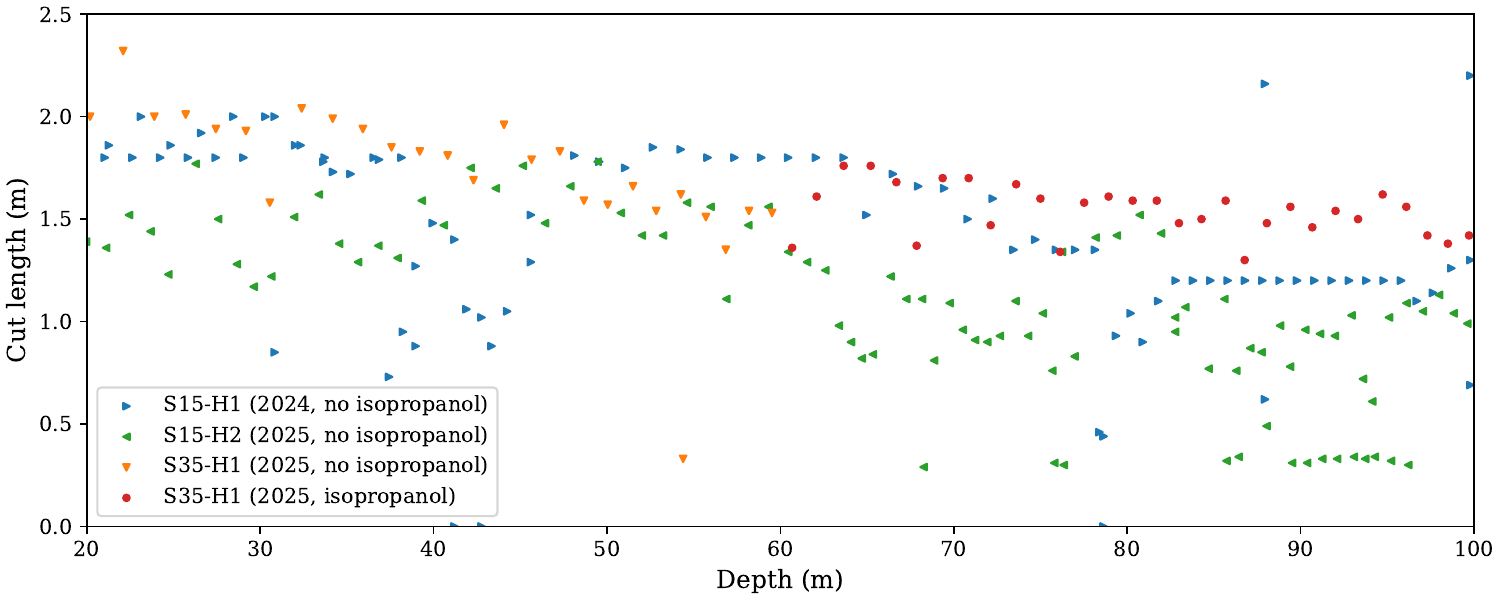}}
\caption{Comparison of cut length vs depth for a hole drilled in 2024 (Station 15 Hole 1), a hole drilled in 2025 (Station 15 Hole 2) and a hole drilled in 2025 (Station 35 Hole 1). The first two holes shown here were drilled without using any isopropanol; for the third hole, isopropanol was applied at every drilling run below 60\,m depth. In all the three holes staggered insert cutters were used.}
\label{fig:isopropanol}
\end{figure*}

A new and not yet fully understood drilling issue also emerged during the 2025 season at depths below approximately 50\,m. Under these conditions, the drill would occasionally fail to engage the ice and begin cutting. When retrieved to the surface and redeployed without modification or cleaning, the sonde would typically re-engage but produce cuts of only 0.8\,m or less. No evidence of ice build up on the cutters was observed after these empty runs. Replacing cutters and performing more extensive cleaning procedures did not alter this behavior. The team subsequently began spraying a minimal amount of isopropanol on the cutter head prior to every drilling run, after which the issue disappeared and cut lengths increased substantially. This behavior is illustrated in Fig.~\ref{fig:isopropanol}, which compares cut length versus depth for a hole drilled in 2024, an early-season 2025 hole, and a late-season 2025 hole.

\section{Upcoming improvements}
In this section, additional improvements to the system that will be tested during summer 2026 are presented. 

\subsection{Drill assembly table}
\begin{figure*}[h]
    \centering
    \includegraphics[width=1\linewidth]{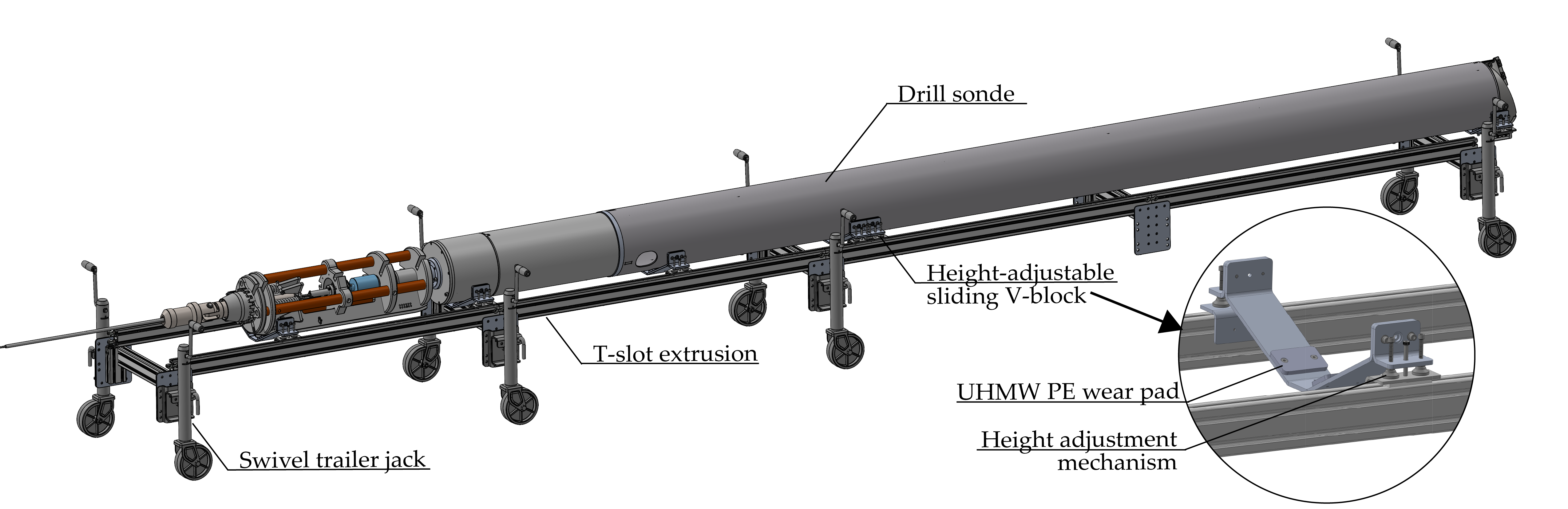}
    \caption{CAD rendering of the modular drill assembly table system. The system consists of T-slot aluminum extrusions, sliding V-block brackets, and adjustable swivel jacks.}
    \label{fig:assembly_table}
\end{figure*}

A modular 6\,m long assembly table was designed and fabricated to streamline the drill assembly process. The assembly table, as shown in Fig. \ref{fig:assembly_table} consists of four 1.5\,m sections of T-slot extrusions (\emph{80/20 15 Series}) that can be combined in various lengths and configurations using custom aluminum plates and brackets. The table is supported by swivel jacks that are typically used on automotive trailers. These jacks allow for height adjustment of the table surface and for tilting the drill sonde to evacuate melting snow and ice. Sonde segments are supported on the table using sliding V-blocks made of custom bent aluminum sheet metal with acetal and ultra high molecular weight polyethylene (UHMW PE) wear pads. The V-blocks incorporate a centerline height adjustment mechanism to concentrically align sonde segments of slightly different diameters. Since the segments were designed with tight dimensional tolerances and fits, this ability to methodically and finely adjust the centerline height will allow for a smoother assembly process. The height-adjustable legs of the table will also provide a more ergonomically-friendly workspace for drill operators during the drill assembly.

\subsection{Snowblower traverse system}
Unnecessarily heavy components of the snowblower system were redesigned with lightweight and collapsible replacements. Many of the original snowblower parts were fabricated from thick welded stainless steel. Since these parts are not load-bearing, thinner aluminum counterparts with screw or rivet fastened joints were fabricated. These parts were redesigned with operator ergonomics in mind; handles are integrated and thumb-screws are used. The collapsibility of some of the parts, the hopper, for example, reduces logistical space requirements, and can easily be installed by one person. Riveted joints allow for field-reworking of the parts if needed.

\begin{figure*}[h]
    \centering
    \includegraphics[width=1\linewidth]{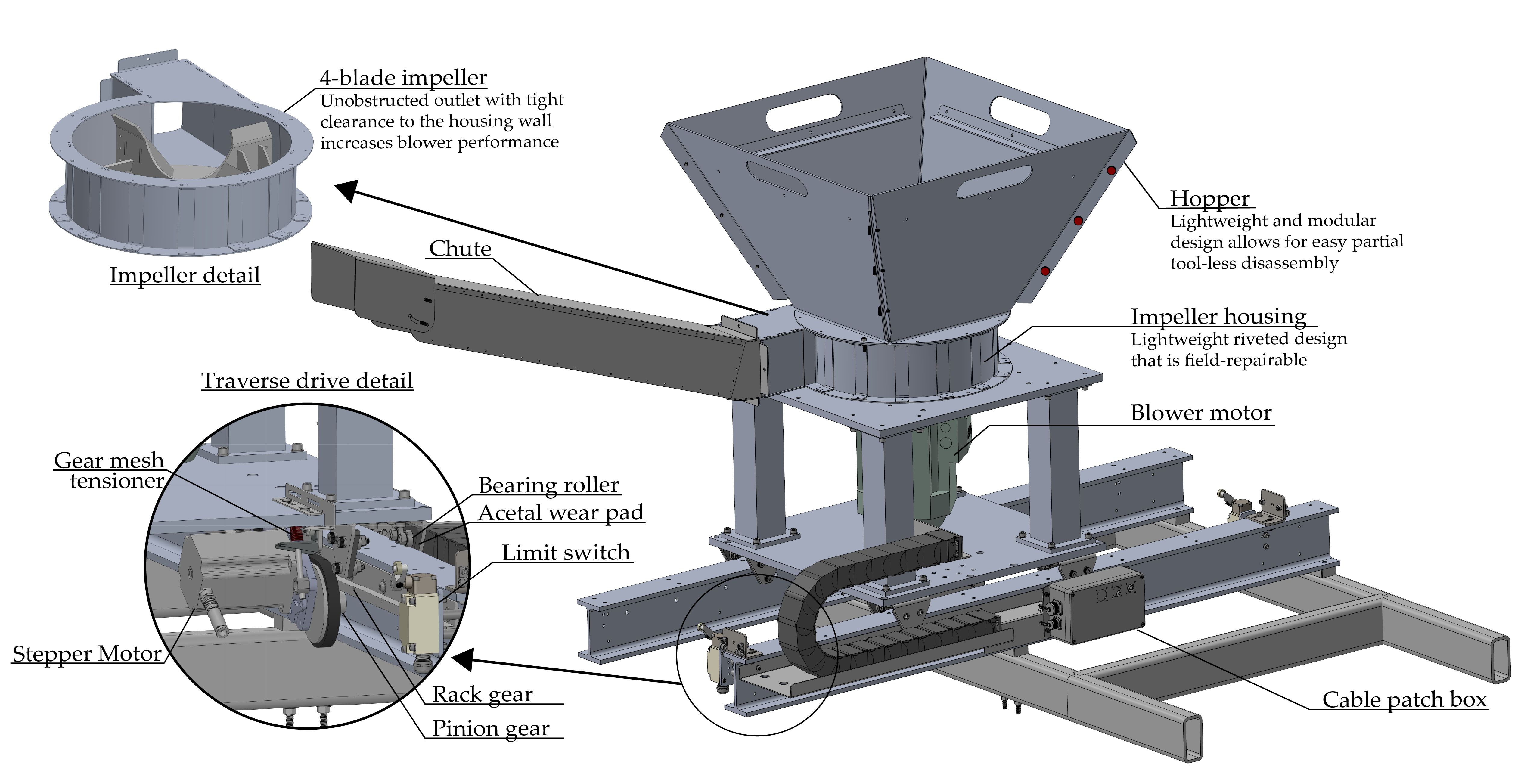}
    \caption{CAD rendering of the redesigned snowblower traverse system used in the 2026 field season.}
    \label{fig:snowblower_2026}
\end{figure*}

\subsection{Drill sonde modifications}
In the last drill season, the redesigned mechanical drill sonde connection proved to work well. However, water ingress near the electrical connection and unreliable electrical grounding needed to be addressed. A more straightforward electrical assembly process was also incorporated in the most recent revision of the drill sonde design. IP65-rated cable glands and a junction box with screw terminals replace a poorly fitting connector. A new electrical slip ring with an extra conductor was fitted to provide a stable ground path while reducing electrical noise by incorporating an extra set of slip ring contacts. 

\subsection{Tent rest kit}
A remaining issue observed during tent raising and lowering operations is the tendency of the tent fabric to become snagged on the tower structure. The edges of the tent are joined using threaded paracord loops, which frequently catch on the pins connecting the individual ST truss sections that form the tower. Attempts to mitigate this problem by taping over the exposed pins to create smoother surfaces provided only partial improvement.

To address this issue for the 2026 season, a new internal guide system has been developed using webbing and line locker rings. The kit will be suspended from the top of the tower inside the tent enclosure and secured above the tower hinges. The purpose of this system is to maintain separation between the tent frame and the tower structure during motion, thereby reducing contact points and minimizing snagging events during deployment and recovery operations. 

\section{Outlook} 
The BigRAID drill system has evolved substantially since its first field deployment in 2021. Over four operational seasons and one dedicated development year, the system transitioned from an early prototype affected by frequent sticking events, control-system instability, and operational inefficiencies into a significantly more robust and mature drilling platform capable of reliably producing deep science boreholes for the RNO-G observatory. \\
A major focus of development has been improving operational reliability under the challenging environmental conditions encountered at Summit Station. Repeated sticking events during the 2021, 2022, and 2024 seasons drove substantial redesign efforts targeting both the mechanical and operational aspects of the sonde. The introduction of a flush cutter-head design in 2025 successfully eliminated sticking events during the entire season, including under abnormal operating conditions such as interruptions in power while actively cutting. This represents a major milestone in the development of the drill.\\
In parallel, extensive improvements were made to the controls and operational infrastructure of the system. The implementation of reliable automated drilling modes, improved subsystem communication, expanded data logging, and additional instrumentation substantially reduced operational overhead and improved situational awareness during drilling. Mechanical upgrades to the cable head, snowblower traverse, and tent handling system further increased operational efficiency and reduced personnel requirements during inter-hole moves. These developments collectively improved both drilling throughput and field maintainability. \\
Despite the significant progress achieved to date, several challenges remain. Temperature-related issues continue to affect some subsystems, particularly the snow-ejection system and barrel icing during transport between holes. In addition, a previously unobserved drilling behavior emerged during the 2025 season at depths below approximately 50--70\,m, where the drill intermittently failed to engage the ice effectively. Preliminary mitigation using isopropanol resulted in substantial improvement, but the underlying mechanism remains unclear and requires further investigation.\\
Future work will focus on improving reliability and automation while reducing operational complexity and crew requirements. The current version of the cutter head is equipped for shoe sockets which should limit the cutting pitch therefore reduce current spikes and increasing cut lengths, however the combination of shoes and shims tried in the 2025 season didn't show significant improvements and will be subject of further investigation this upcoming season. Further integration of sensor data and automated monitoring may also enable predictive diagnostics and improved autonomous operation. As RNO-G deployment progresses toward the full array, continued refinement of the BigRAID platform will remain essential to achieving efficient and repeatable borehole production in remote polar environments.

\section{Acknowledgments} 
We are thankful to the support staff at Summit Station for making RNO-G possible. We also acknowledge our colleagues from the British Antarctic Survey for building and operating the BigRAID drill for our project.

We would like to acknowledge our home institutions and funding agencies for supporting the RNO-G work. In particular, the Belgian Funds for Scientific Research (FRS-FNRS and FWO), the FWO programme for International Research Infrastructure (IRI), the National Science Foundation (NSF Award IDs 2411590, 2514262, 2514201, 2514208, 2514206, and collaborative awards 2310122 through 2310129), the IceCube EPSCoR Initiative (Award ID 2019597), the Helmholtz Association, the German Research Foundation (DFG, Grant 556092823), the Swedish Research Council (VR, Grant 2023-00156), the University of Chicago Research Computing Center, and the European Union under the European Unions Horizon 2020 research and innovation programme (ERC, Pro-RNO-G No 101115122 and NuRadioOpt No 101116890) made this work possible.

\bibliography{Bibliography}   
\bibliographystyle{igs}  

\end{document}